\documentclass[fleqn,usenatbib]{mnras}

\usepackage{newtxtext,newtxmath}

\usepackage[T1]{fontenc}

\DeclareRobustCommand{\VAN}[3]{#2}
\let\VANthebibliography\thebibliography
\def\thebibliography{\DeclareRobustCommand{\VAN}[3]{##3}\VANthebibliography}

\usepackage{graphicx}	
\usepackage{amsmath}	

\def\vel{{\rm v}}
\def\velvec{{\rm \mathbf v}}
\def\Rstar{{\rm R_{\rm star}}}

\newcommand{\AvgS}[1]{\left< {#1} \right>_{\mathcal{S}}}
\newcommand{\AvgST}[1]{\left< {#1} \right>_{\mathcal{S},t}}
\newcommand{\AvgT}[1]{\left< {#1} \right>_t}

\title[Amplitudes of p modes in rotating solar-like stars]{Impact of rotation on the amplitude of acoustic modes in solar-like stars: Insights from hydrodynamical simulations}

\author[A. Le Saux et al.]{
Arthur Le Saux,$^{1}$\thanks{E-mail: arthur.lesaux@cea.fr}
Leïla Bessila,$^{1,2}$
Stéphane Mathis$^{1}$
\\
$^{1}$Université Paris-Saclay, Université Paris Cité, CEA, CNRS, AIM, Gif-sur-Yvette, F-91191, France.\\
$^{2}$Center for Interdisciplinary Exploration and Research in Astrophysics (CIERA), Northwestern University, 1800 Sherman Ave, Evanston, IL 60201, USA
}

\date{Accepted XXX. Received YYY; in original form ZZZ}

\pubyear{\the\year{}}

\begin{document}
\label{firstpage}
\pagerange{\pageref{firstpage}--\pageref{lastpage}}
\maketitle

\begin{abstract}
In solar-like stars, acoustic modes provide the main way of probing their internal structure and dynamics. Although these modes are expected to be ubiquitous in stars with convective envelopes, \textit{Kepler} observations reveal that a significant fraction of solar-like stars show no detectable acoustic modes, particularly among rapidly rotating and magnetically active stars.
Recent theoretical work has proposed that rotation tends to inhibit convective motions, thereby reducing the power available for stochastic excitationof low degree acoustic modes. Here, we test this prediction using fully compressible hydrodynamical simulations of a solar-like star. We perform a series of 2.5D simulations, which consider longitudinal symmetry, using the \textsc{MUSIC} code spanning rotation rates from 0 to 8 $\Omega_{\odot}$. We find a clear and systematic decline of acoustic mode amplitudes with increasing rotation rate. In the most rapidly rotating models, mode damping rates are also enhanced. 
The combined reduction in excitation and increase in damping with increasing rotation rate provide a physical explanation for the observed decrease in mode detectability in rapidly rotating solar-like stars. Our results demonstrate that rotation can significantly modify oscillation properties and must be accounted for when interpreting asteroseismic observations.

\end{abstract}




\section{Introduction}

The era of high precision photometric space-based missions has revolutionised our understanding of stars. First, CoRoT \citep{Auvergne2009}, then \textit{Kepler}/K2 \citep{Borucki2010, Howell2014} and now TESS \citep{Ricker2014} have provided high quality data that allowed detecting global oscillation modes in hundreds of thousands of stars. Most of these oscillations are acoustic modes, resulting from the interferences of sound waves that propagate in the interiors of stars and make their surfaces flicker.
Such kinds of oscillations were first detected in the Sun with pioneering studies by \cite{Ulrich1970} and \cite{Leibacher1971} that identified the solar five-minute oscillations as global acoustic standing waves, known as $p$ modes for pressure modes.
Thus, when similar oscillations were observed in other stars, it led to the denomination of the large stellar class of solar-like oscillators. 
Since the launch of CoRoT in 2006, there have been catalogs of hundreds of thousands of solar-like oscillators \citep[e.g.][]{Chaplin2014, Mathur2016, Mathur2022, Yu2016, Yu2018, Hon2018, Hon2019, Santos2019, Hatt2023, Zhou2024}. For these stars, acoustic modes have been crucial in determining their global parameters, such as their mass, age and radii, as well as probing the physical processes taking place in their interior \citep[see e.g. the reviews of][ and references therein]{Aerts_AngularMomentumTransportStellarInteriors_2019, Garcia2019}.

Solar-like oscillations are driven by turbulent convective motions in the envelope of a star \citep{Goldreich1977}. Thus, it groups stars, on the main-sequence or post-main-sequence, that are cool enough to have such a convective envelope. More precisely, it is the Reynolds stresses generated by these turbulent motions that inject some power into acoustic modes. This mechanism is often referred to as stochastic excitation \citep{Balmforth1992, Goldreich1994, Samadi2001}. Consequently, for every star that possesses a convective envelope it is expected to detect $p$-modes. 
However, through the analysis of late-type stars observed by the \textit{Kepler} telescope, \citet{Chaplin2011} and \citet{Mathur2019} discovered that a significant fraction of these stars, which do have a convective envelope, do not exhibit solar-like oscillations. \cite{Mathur2019} actually detected acoustic modes in only 46\% of the stars in their sample. The authors found that this non-detection is correlated with rotation and magnetic field, with almost no $p$ modes detected for stars with a rotation period smaller than five days neither for magnetically active stars.

Recent theoretical work by \citet{Bessila2024} has proposed that rotation can inhibit the excitation of acoustic modes by reducing the power injected into the modes through its indirect impact on convection. Indeed, the influence of rotation on convection has long been recognised. \citet{Chandrasekhar1961} already demonstrated that rotation has a stabilising effect, making the convective instability more difficult to trigger. More recent dedicated numerical studies have confirmed that turbulent convection is strongly impacted by rotation, which alters its morphology, anisotropy, and transport properties \citep[e.g.][]{Brun2017, Brun2022, Hindman2020}.
In stellar evolution models, convection is most of the time modelled with the Mixing-Length Theory \citep[MLT, ][]{Bohm-Vitense1958}, a monomodal approach that focuses on the convective mode that carries the most heat. Such model can be extended to include the effect of rotation through the Rotating Mixing-Length Theory \citep[R-MLT, ][]{Stevenson1979, Augustson2019}, which has been shown to give good agreement with non-linear hydrodynamical simulations in a localised Cartesian framework \citep[e.g.][]{Barker_Theorysimulationsrotatingconvection_2014, Currie2020}, global simulations focusing on the overshoot region \citep[e.g.][]{Korre2021} and observations \citep[e.g.][]{Dumont2021,Michielsen2019}. Building on existing models of stochastic excitation and incorporating the effects of rotation through the R-MLT, \citet{Bessila2024} showed that uniform rotation tends to reduce the power injected into acoustic oscillations. In a solar-like star rotating at $20 \Omega_{\odot}$, they predict a decrease up to 70\%. This study was further extended by \citet{Biscarrat2025} to the differentially rotating case, showing that it can significantly increase (resp. decrease) the power injected into the modes in the case of solar (resp. anti-solar) conical differential rotation compared to the uniformly rotating case. Solar differential rotation refers to an equatorial region rotating faster than the poles, and it is the opposite for anti-solar.

In this Letter, we confront the theoretical predictions of \citet{Bessila2024} using dedicated numerical simulations, with particular emphasis on the impact of rotation on the excitation of acoustic modes. Our simulations also enable us to characterise the damping of these modes and thereby quantify the influence of rotation on their global amplitudes. To date, multidimensional simulations of waves in solar-like stars have primarily focused on internal gravity waves \citep[e.g.][]{Rogers2006,Alvan2014,Breton2022}. This emphasis is largely due to the codes used, which are based on the anelastic approximation and thus filter out acoustic waves.
Although $p$ modes have been identified in global simulations of massive stars \citep[e.g.][]{Meakin2006,Horst2020,Thompson2024}, their properties were not analysed in detail, as those studies concentrated on internal gravity waves. Only a limited number of investigations have specifically addressed acoustic modes, and these were restricted to local simulations of the near-surface stellar layers and to radial modes \citep{Stein2001,Belkacem2019, Zhou2019, Zhou2020}. Here, we report the first dedicated analysis of solar-like oscillations in global rotating stellar simulations. 
In Sec. \ref{sec:theory}, we summarise the main outputs from the theory of \cite{Bessila2024}. In Sec. \ref{sec:results}, we present hydrodynamical simulations of a $1 M_{\odot}$ star similar to the Sun with five different rotation rates $\Omega = 0, 1$, 3, 5 and 8$\Omega_{\odot}$. Finally, we conclude in Sec. \ref{sec:conclusion} with a summary of our findings and perspectives to come.

\section{Theory for the stochastic excitation in rotating solar-type stars}
\label{sec:theory}

In this section, we recall the main results to compute the power injected into the stochastically-excited acoustic modes. We refer the reader to \cite{Bessila2024} for more details regarding the derivation. In this work, the centrifugal force and magnetic effects are neglected \citep[see][for the impact of magnetic field on the stochastic excitation of solar-like oscillations]{Bessila2024a}, thus it is assumed as a first step that only the Coriolis force impacts the excitation. However, the source term that emerges for the Coriolis force is negligible in solar-like stars \citep[see e.g.][]{Belkacem2009b}. In that case, the dominant source driving the oscillations comes from the Reynolds stresses related to the turbulent velocities, and the power injected into a given acoustic mode writes \citep{Samadi2001, Belkacem2009b}: 
\begin{equation}
    \mathcal{P} = \frac{C_R}{8I},
\label{eq:power}
\end{equation}
where $I$ is the mode inertia and $C_R$ is the contribution of the Reynolds-stresses source term. Following \citet{Samadi2001}, we consider that the wavelength of oscillations is large compared to typical length scale of turbulence (separation of scales) and that turbulence is homogeneous and isotropic to compute $C_R$. It is assumed that the plane-parallel approximation is valid, which is the case when $r k_{\rm osc} \gg 1$, with $k_{\rm osc}$ the local wavenumber of a given acoustic mode \citep{Belkacem2008}. We also assume that acoustic modes are mostly radial i.e. $\xi_r \gg \xi_{\rm h}$, with $\xi_r$ and $\xi_{\rm h}$ the radial and horizontal functions of the displacement eigenfunction, respectively.
In this framework, we obtain: 
\begin{equation}
    C_R=\frac{16}{15} \pi^3 \int_{\mathcal{V_{CZ}}} d^3 x_0 \rho_0^2\left|\frac{\mathrm{d} \xi_r(r)} {\mathrm{~d} r}\right|^2 Y_{\ell,m}(\theta,\varphi) Y^{*}_{\ell,m}(\theta,\varphi)\hat{S}_R\left(r,\theta,\omega_0\right) ,
    \label{eq:reynolds-int}
\end{equation}
where $\rho_0$ is the density, $Y_{\ell,m}$ are the usual spherical harmonics, and $Y_{\ell,m}^{*}$ their complex conjugates. The spherical harmonics are characterised with an angular degree $\ell$ and an azimuthal order $m$. We use here the spherical coordinates with radius $r$, co-latitude $\theta$ and longitude $\varphi$. The integration is performed on the whole volume of the convective zone $\mathcal{V}_{CZ}$. The term $\hat{S}$ is the spectral source, which is written: 
\begin{equation}
\hat{S}_R\left(r, \theta, \omega_0\right)=\int_{k_c}^{+\infty} \frac{d k}{k^2} E^2(k) \mathcal{I}(\omega_0,k), 
\label{eq:s_hat}
\end{equation}
where $k$ is the wavenumber corresponding to a given eddy in the turbulent cascade, $E(k)$ is the spatial part of the kinetic energy spectrum. 
Considering isotropic turbulence, and a Kolmogorov-like spectrum \citep{Kolmogorov1941}, we have $E(k) \propto k^{-\alpha}$. Usually, $\alpha = 5/3$, but \citet{Bessila2024} have also tested the influence of choosing $\alpha = 2$ and $3$ as found for rotating turbulence in numerical simulations \citep{Zhou1998, Smith1999}, and showed that the slope of the turbulent spectrum does not change the overall tendency.
In addition, we have: 
 \begin{equation}
    \mathcal{I}(\omega_0, k) \equiv \int_{-\infty}^{+\infty} \chi_k(\omega) \chi_k(\omega + \omega_0) d \omega,
    \label{eq:i_omega} 
\end{equation}
where $\chi_k$ is the eddy-time correlation function as introduced in \citet{Stein1967}. Two types of functions are usually assumed for this correlation : Gaussian or Lorentzian (see Appendix \ref{app:eddy}). An important result from \citet{Bessila2024} is that the choice of this correlation function is crucial to model the excitation of $p$ modes. When choosing a Lorentzian eddy-time correlation function, the stochastic excitation is significantly impacted by rotation, whereas it is not when choosing a Gaussian function. The reason of this behaviour is that the Gaussian correlation function neglects the largest scales when estimating the power injected into $p$ modes, while these large scales are those, which are the most impacted by the Coriolis acceleration \citep[see Fig. 4 and Sect. 4.1 in][for more details]{Bessila2024}. As in observations acoustic mode detection seems to depend on the rotation period \citep{Mathur2019}, we make the choice of a Lorentzian (see Eq. \ref{eq:chi_l}). 
In this framework, \citet{Bessila2024} predicts that rotation can significantly reduce the power injected into $p$ modes. We shall now confront these results with non-linear hydrodynamical simulations.

\section{Results from numerical simulations}
\label{sec:results}

\begin{figure*}
   \resizebox{\hsize}{!}
            {\includegraphics[]{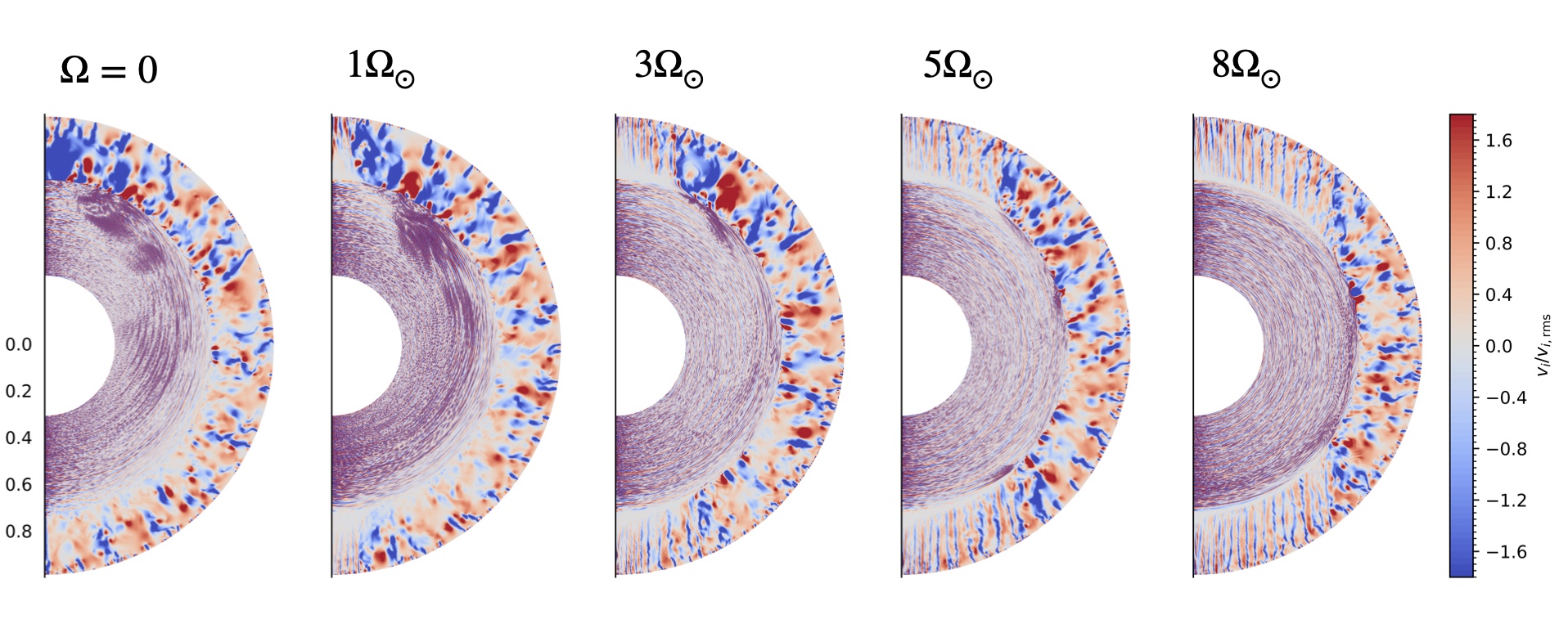}}
      \caption{Snapshots of the instantaneous radial velocity $\vel_r$ in the five simulations, normalised by its root-mean-square value at each radius. Red indicates outward and blue inward motions.
              }
         \label{fig:vel1}
\end{figure*}

We have run five hydrodynamical simulations using the \textsc{music} code \citep{Viallet2013,  Viallet2016, Goffrey17}. This code solves the fully compressible equations of hydrodynamics; thus, it is particularly adapted for asteroseismic studies, as it can model $g$, $f$ and $p$ modes, as well as their coupling \citep[see][for the coupling of $f$ and $g$ modes]{LeSaux2025}. 
The 1D stellar evolution model used to initialize the simulations corresponds to a 1 M$_{\odot}$ star of 4.6 Gyrs old with a radius $R_{\rm star}$ = 1.05 R$_{\odot}$ (with $R_{\odot}= 6.957\times 10^{10}$ cm the solar radius). The base of the convective envelope is located at $r$ = 0.722 $R_{\rm star}$. It has been constructed using the \textsc{MESA} stellar evolution code \citep{Paxton2011, Paxton2013, paxton2015, paxton2018, paxton2019, jermyn2023} and is the same as the main-sequence stellar model used in \citet{Bessila2024}. For details on the set-up of simulations, we refer the reader to Appendix \ref{app:sims}.
The five simulations only differ in rotation rates, which are $\Omega $ = 0, 1, 3, 5 and 8 $\Omega_{\odot}$. For the reference solar rotation rate, we use $\Omega_{\odot}$ = 413 nHz \citep{Thompson2003, Garcia2007}. The five simulations cover a spatial domain $r \in $ [0.3; 0.98] $R_{\rm star}$ and $\theta \in  [0; \pi]$. These are 2.5D simulations, a hybrid approach combining the advantages of 3D simulations - which are required to include rotation - but considering azimuthal symmetry and thus reducing their numerical cost to that of a 2D simulation to explore the parameter space. We then have three components for the velocities that depend only on two coordinates, that is, $\vel_i(r,\theta)$ with $i =$ \{r, $\theta$, $\varphi$\}. Similarly, all variables depend on these two coordinates.

\subsection{Velocities in the convective zone}
Figure \ref{fig:vel1} presents instantaneous snapshots of the radial velocity in the five simulations. The velocity is normalised by its root-mean-square (rms, defined in Appendix \ref{app:rms}) at each radius for better visibility, as its magnitude in the radiative region is a few orders of magnitude smaller than in the convective zone (CZ). The distinction between the radiative interior and the convective envelope can be clearly seen from the different nature of fluid motions in the two zones. In the envelope, the aspect of convective motions changes with varying rotation rate. 

In the non-rotating case (left panel), convection is dominated by large-scale radial upflows and downflows spanning the depth of the convective envelope. At the solar rotation rate (second panel from the left), the global morphology remains broadly similar throughout most of the convection zone, although thinner structures begin to emerge near the poles. In that case, convection is dominated by advective forces. This can be confirmed by computing the Rossby number $Ro$, which represents the relative importance of advective to Coriolis forces, and is defined in Eq. \eqref{eq:rossby}. We find that $3 > Ro \geq 2$ in the whole CZ at $\Omega= 1 \Omega_{\odot}$. We refer the reader to Appendix \ref{app:rossby} for more details. 
As the rotation rate increases (i.e. as the Rossby number decreases), rotational constraints become progressively dominant: convective motions align with the rotation axis and the flow approaches a quasi-cylindrical regime consistent with the Taylor–Proudman theorem. For the 8 solar rotation rate case (right panel), these columns extend from the pole to mid-latitudes, and we find that $Ro < 1$ at these high latitudes, whereas $Ro \sim 1$ near the equator.
Simultaneously, the characteristic convective length scale decreases, reflecting the suppression of large-scale motions by the Coriolis force. This trend is in agreement with previous simulations of rotating solar convection \citet{Brown2008}, which report a decrease in horizontal convective scales and enhanced alignment with the rotation axis as the Rossby number decreases.

In addition, the magnitude of the convective velocities is also impacted by rotation. As the rotation rate increases, the rms velocity decreases in the CZ. This is particularly obvious in the bottom of the convection zone, near the convective boundary (see Fig. \ref{fig:vrms}). This decrease of the magnitude of convective velocities with rotation rate follows the trend predicted by R-MLT \citep[][]{Stevenson1979, Bessila2025} as illustrated in Fig. \ref{fig:RMLT}. 
However, this trend does not seem to hold in the non-rotating case. We show that this is an effect of the 2.5D configuration, and we are planning a follow-up analysis to compare 2.5D and 3D simulations to confirm this.


\subsection{Amplitudes of acoustic modes}
\label{sec:pmodes_ampl}

\begin{figure*}
   \resizebox{\hsize}{!}
            {\includegraphics[]{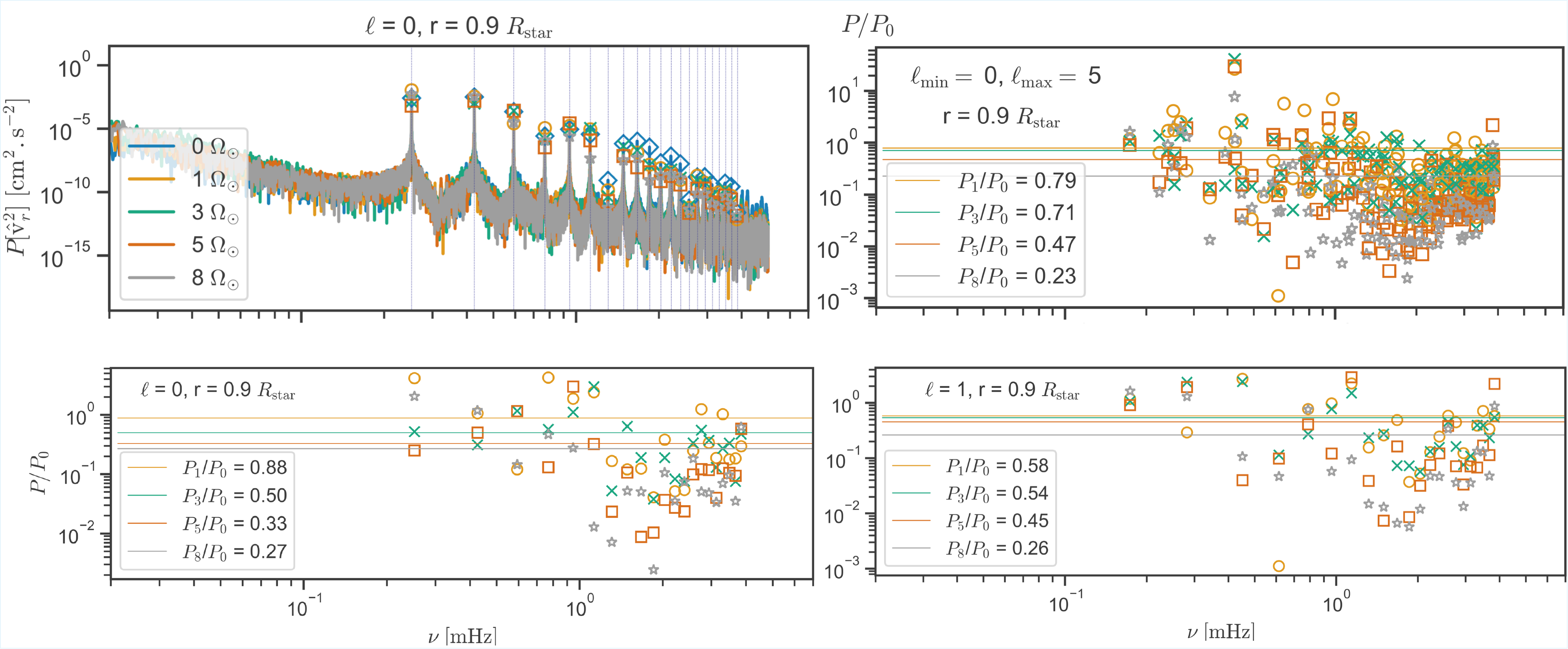}}
      \caption{Comparison of the power spectra of the radial velocity measured in the simulations at $r=0.9 R_{\rm star}$ for angular degree $\ell$ = 0 (top left panel). We also show the eigenfrequencies of the 1D model predicted by GYRE (vertical blue dashed lines).  
      The ratio of the  $p$ modes' amplitudes in a rotating simulation to the non-rotating case for $\ell \in [0;5]$ (top right), $\ell  = 0$ (bottom left) and $\ell = 1$ (bottom right) as a function of frequency for 5 different rotation rates. The horizontal lines represent the amplitude ratios averaged over all modes of a given simulation.
              }
         \label{fig:comp_ellrange}
\end{figure*}

In this section, we identify $p$ modes in our five simulations, then measure and compare their amplitudes. As $p$ modes are mostly radial perturbations, we do so by computing the power spectra of the radial velocity $P[\hat{\rm v}_r^2](r,\ell,\nu)$ measured in the simulations, with $\nu$ the frequency. These power spectra are computed with the method described in detail in \cite{LeSaux2022, LeSaux2023}.
In the top left panel of Fig. \ref{fig:comp_ellrange}, we show these power spectra at fixed angular degree $\ell = 0$ to 5, and at the depth r = 0.9$R_{\rm star}$ for the five rotation rate $\Omega $ = 0 (blue), 1 $\Omega_{\odot}$ (orange), 3 $\Omega_{\odot}$ (green), 5 $\Omega_{\odot}$ (red) and 8 $\Omega_{\odot}$ (grey). 
The acoustic modes in the simulation are the narrow high amplitude peaks in the power spectra and are identified using the eigenfrequencies (vertical dotted dark blue lines) of the initial 1D MESA model computed with the GYRE  oscillations code \citep{townsend2013, townsend2018}. Using this identification, we measure the corresponding amplitude of the modes and indicate it as coloured symbols : rhombus, circle, cross, square and star for $\Omega $ = 0, 1, 3, 5 and 8 $\Omega_{\odot}$ respectively, with the same colour code as for the curves.

Next, we select all $p$ modes whose frequency lie between the Lamb frequency $L_{\ell} = \frac{\sqrt{\ell(\ell+1)}}{r}c_s$ (dotted black line), with $c_s$ the sound velocity, and 4 mHz. Indeed, a necessary condition for an acoustic wave to propagate is that its frequency $\nu$ is larger than $L_{\ell}$ and we do not include too high frequency modes as these modes should propagate in the uppermost layer of the Sun, which are not modelled here.

In the bottom left panel of Fig. \ref{fig:comp_ellrange}, we compare the ratios of the modes' amplitude in a rotating case $P_i$, with $i$ = 1, 3, 5, 8, to the non-rotating case $P_0$. The symbols are the same as for the upper left panel. We then compute the average amplitude ratios (horizontal solid lines with corresponding color code) using all modes for each simulation.
In the $\Omega = 1 \Omega_{\odot}$ simulation, we measure an average decrease in amplitude  of 21\%. In the $3 \Omega_{\odot}$ case, this decrease is of 29\%, in the $5 \Omega_{\odot}$ case of 53\% and in the $8 \Omega_{\odot}$ case of 77\%. 
We have also computed these ratios in the radiative zone at $r = 0.6 R_{\rm star}$ and for angular degrees $\ell$ between 0 and 20 (see Fig. \ref{fig:comp_largeL} in Appendix \ref{app:extra_ampl}), and get a similar trend.
The spectra were computed using the last $5 \times 10^6$ s of each simulation as well as for the same duration but $\sim 5 \times 10^7$ s earlier. We do not find any significant variation of the average amplitude, meaning that the simulations are converged with respect to the amplitudes of the modes.

\begin{figure}
   \resizebox{\hsize}{!}
            {\includegraphics[]{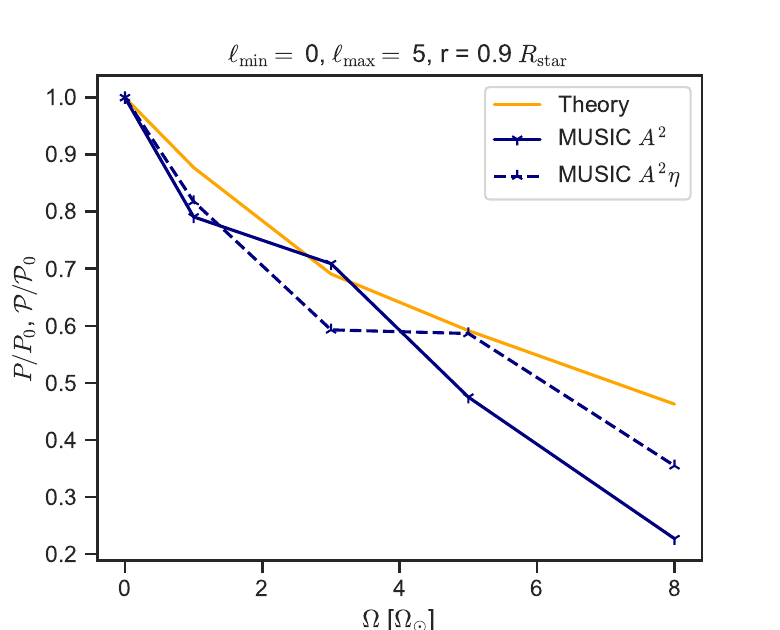}}
      \caption{Variation with the rotation rate of the ratio of $p$ modes' amplitude $P$ measured in rotating simulations to the amplitude in the non-rotating simulation $P_0$ (plain blue curves), and of the ratio of the power injected into the modes in the rotating to the non-rotating case, predicted theoretically (orange curve) and measured in simulations (dashed blue curve).
              }
         \label{fig:main_comp}
\end{figure}

Although the decrease in amplitude of $p$ modes with rotation is clear, we can notice that the absolute value of the decrease can vary significantly with angular degree and frequency. The right panels of Fig. \ref{fig:comp_ellrange} present the amplitude ratio measurements but for only one given angular degree $\ell = 0$ (top) and $\ell = 1$ (bottom). We find that for a given angular degree the change in amplitude ratio for distinct modes can vary by more than 2 orders of magnitude. 
This might be expected, as stochastic convective motions are exciting modes that are randomly triggered. For this reason, in order to compare our results with the theoretical predictions of \citet{Bessila2024} we use the amplitude ratios averaged over all modes between angular degrees $\ell = 0$ and 5. This is presented on Fig. \ref{fig:main_comp}. The dark blue curve reports the amplitude ratios $P_i/P_0$ of Fig. \ref{fig:comp_ellrange} bottom left panel. The orange curve shows the theoretical predictions for the ratios of power injected into the modes $\mathcal{P}_i / \mathcal{P}_0$, with $\mathcal{P}$ given in Eq. \eqref{eq:power} and computed using the method presented in \citet{Bessila2024}. The power is averaged over eigenmodes of the MESA model between $\ell = 0$ and 5, and for radial orders $n$ = \{4, 7, 10, 13, 16, 19, 22, 25, 28\}.
In addition, we validate thanks to our hydrodynamical simulations the choice of \citet{Bessila2024} to use a Lorentzian function for the eddy-time correlation. To do so, we compute the eddy-time correlation functions for convective motions in our five simulations and find that they are well approximated by Lorentzian functions in the $p$ modes frequency range (we refer the reader to Appendix \ref{app:eddy} for details).
As a result, we can see that for rotation rate between 1 and 5 $\Omega_{\odot}$ the decrease in amplitude measured in simulations follows approximately the trend predicted theoretically for the decrease in power injected into the mode. For the fastest rotating simulations, the decay in amplitude is more important than the one predicted by the theory.

\subsection{Damping of $p$ modes}
\label{sec:mode_damping}
To investigate this discrepancy, we look into the damping of acoustic modes by convective and radiative effects. Indeed, the actual amplitude of a mode is set by a balance between excitation and damping and the theoretical work of \citet{Bessila2024} only accounts for the impact of rotation on the excitation of the modes. With our non-linear simulations, we can study the nonadiabatic effects responsible for mode damping and quantify the impact of rotation on it. 

The methodology to measure the damping rate $\eta$ of a mode is described in detail in Appendix \ref{apdx:damping}. We give here an overview of the main idea. The damping rate $\eta$ of a mode with amplitude $A$ can be directly inferred from the simulations by measuring the linewidth $2\Gamma$ at mid-amplitude $A/2$ of the peak of the corresponding mode in the power spectrum of the radial velocity. The relation between the linewidth and the damping rate is given by \citep{Belkacem2019}:
\begin{equation}
    \Gamma = \frac{\eta}{2\pi}.
    \label{eq:damping}
\end{equation}
To measure $\Gamma$, we use the normalised the power spectrum of the radial velocity $P_{\rm norm}[\hat{\rm v}_r^2]$, whose maximal value is thus 1, and fit a Lorentzian profile to each peak corresponding to a $p$ mode. This is illustrated in Fig. \ref{fig:damping_l1_main} for a $p$ mode with $\ell = 4$ and $\nu = 2.243$ mHz. As explained in Appendix \ref{apdx:damping}, we only include modes for which $\Gamma > 0.4$ $\mu$Hz to avoid contamination from the window function used to compute the power spectra.
We then obtain $\eta_0 = 26.84$ $\mu$Hz, $\eta_1 = 27.76$ $\mu$Hz, $\eta_3 = 22.45$  $\mu$Hz, $\eta_5 = 33.15$  $\mu$Hz and $\eta_8 = 40.59$  $\mu$Hz for the simulations with $\Omega$ = 0, 1, 3, 5 and 8 $\Omega_{\odot}$, respectively. This represents variations in damping rates of 3\%, -16\%, 24\% and 51\% for the simulations with $\Omega$ = 1, 3, 5 and 8 $\Omega_{\odot}$, respectively. The minus sign for the $\Omega = 3 \Omega_{\odot}$ case indicates that the damping is a bit weaker than in the non-rotating case.

While there is not a clear trend with rotation, these measurements nevertheless suggest that the damping of $p$ modes seems to be larger for simulations with higher rotation rate. As detailed in Appendix \ref{apdx:damping}, this result has to be taken with caution as we only consider modes for which we can clearly identify the peak in the power spectrum.
Now that we have estimations of the damping rates of the modes, we can calculate the power injected into the modes in the simulations using $\mathcal{P} \sim P[\hat{\rm v}_r^2] \eta$ (see Eq. \ref{eq:Pinj_sims}).
This quantity is plotted as the dashed blue curve in Fig. \ref{fig:main_comp}. The values of the ratio at a given rotation rate are a bit shifted compared to the one measured using only the amplitudes of the modes. Even if this shift can reach up to $\sim 10\%$, the power injected into modes is still significantly reduced as the rotation rate increases. 

\begin{figure}
   \resizebox{\hsize}{!}
            {\includegraphics[]{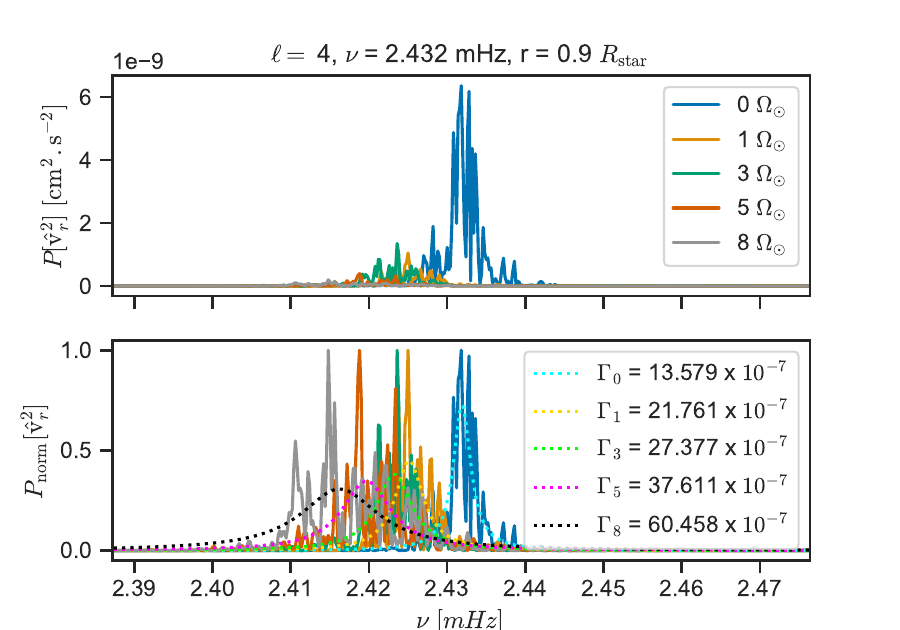}}
      \caption{Measurement of the half linewidth $\Gamma$ of the $p$ mode $\ell = 4$ and  $\nu = 2.243$ mHz. The upper plot shows the power spectra $P[\hat{\rm v}_r^2]$ computed using the radial velocity in each simulation, and the lower plot shows the normalised power spectrum $P_{\rm norm}[\hat{\rm v}_r^2]$ (plain curves) and the Lorentzian fit (dotted lines). 
              }
         \label{fig:damping_l1_main}
\end{figure}

We have also computed the theoretical estimation of the power injected into the modes using Eq. \eqref{eq:power} but this time using the kinetic energy spectra and eddy-time correlation function measured in the simulations instead of using analytical prescriptions. While the injected power does not reach its maximum in the non-rotating case, consistent with the rms velocity measurements, we still measure a decrease in injected power for more rapidly rotating stars. We refer the reader to Appendix \ref{app:Pinj_sims_spectra} for more details.
Overall, all estimates presented in Fig. \ref{fig:main_comp} exhibit a consistent dependence on rotation rate, reinforcing the robustness of this result.

\section{Discussion \& Conclusions}
\label{sec:conclusion}
Our results confirm the significant impact of rotation on the amplitude of $p$ modes in solar-like stars. Using non-linear global hydrodynamical simulations of a 1M$_{\odot}$ star, we measure a decrease in the amplitude of low degree $p$ modes of approximately 21\%, 29\%, 53\% and 77\% for stars rotating at a rate $\Omega = 1, 3, 5$ and $8 \Omega_{\odot}$, respectively, with respect to a non-rotating model. 
We have also studied the damping of acoustic modes in our simulations and our results suggest a correlation with rotation, with a damping rate increased by almost 50\% for the fastest rotating case. However, a dedicated study is required to confirm this trend, as we have not considered the lowest frequency modes (for which the mode linewidth is smaller than our frequency resolution) and the highest frequency (which propagates in the outermost layers of the star which are not included in our numerical models).
To enable a direct comparison with the theoretical predictions of \citet{Bessila2024}, we use the measured damping rate to estimate the power injected into the modes through the excitation by convection. We find that the injected power decreases by 15\%, 28\%, 41\% and 73\% for stars rotating at a rate $\Omega = 1, 3, 5$ and $8 \Omega_{\odot}$, respectively.
These values agree relatively well with the theoretical predictions, which predict a reduction of approximately $\sim$13\%, $\sim$30\% $\sim$40\% and $\sim$55\% for low-degree modes (0 < $\ell \leq 5$) at rotation rates of $\Omega = 1, 3, 5$ and $8 \Omega_{\odot}$, respectively. Note that in the most rapidly rotating case, the agreement between theory and simulations deteriorates, the simulations exhibiting a stronger decrease than predicted by theory. In a follow-up study, \citet{Biscarrat2025} showed that anti-solar differential rotation can decrease the power injected into the modes by an additional 30\% for a star with a mean rotation of $5 \Omega_{\odot}$. As an anti-solar differential rotation develops in all our simulations, this could explain at least part of the larger decrease measured. We note that, based on the Rossby number measured in the simulations (see Appendix \ref{app:rossby}), particularly in the fastest ones, we should expect a solar-like differential rotation, with an equator rotating faster than the poles \citep[e.g.][]{Brun2017}. This might result from the 2D geometry and will be studied in more details in the analysis of our 3D simulations.

Overall, our results indicate that the reduction in power injected into the modes by convection largely accounts for the decline in $p$ mode amplitudes with increasing rotation rate.
However, when we use the kinetic energy spectra and eddy time-correlation functions measured in our simulations to compute theoretical estimates of the injected power, the maximum does not occur in the non-rotating case. This behaviour is consistent with our measurements of the root-mean-square velocity in the convection zone. This unexpected trend may partly arise from limitations inherent to the 2D numerical configuration, which can influence the structure and energetics of convection. Preliminary analyses of fully 3D simulations indicate higher convective velocities in the non-rotating model than in models rotating at $\Omega = 1$ and $3 \Omega_{\odot}$. However, the reasonable computational cost of 2.5D simulations allows us to explore the required parameter space. A detailed study of dimensional effects will be presented in future work.

While the conclusions we obtain are robust, we emphasize that the quantitative estimates are affected by the usual limitations of numerical simulations. For instance, the boundary conditions which imposes a zero radial velocity on the edge of the domain impact the amplitudes of standing modes. In addition, extending the numerical domain up to r = $R_{\star}$ remains one of the most important challenge of stellar hydrodynamics. This region between r = $0.98 R_{\star}$ and r = $R_{\star}$ will drive stronger convection \citep{Vlaykov2022} and it is the region where the highest frequencies and smaller scales waves are excited. This is the reason why we do not study modes with $\ell > 20$ and $\nu > $ 4 mHz. It is also established that convective velocities are higher in 2D than in 3D simulations \citep[e.g.][]{Pratt2020}. 
These limitations, however, are common to all our simulations. Since our analysis primarily relies on ratios of mode amplitudes, i.e. on relative differences between models, we expect systematic numerical effects to largely cancel out, so that the inferred rotational trends remain robust.

Finally, we show that mode amplitudes, which have been long overlooked in observations, offer promising diagnostics for stellar internal dynamics, complementary to those inferred from modes frequencies. More generally, our results emphasise that a proper understanding of the coupling between stellar internal dynamics and oscillations is essential for reliably interpreting asteroseismic data. This challenge is particularly timely in the view of the upcoming PLATO mission \citep{Rauer2025}, scheduled for launch in December 2026, which will deliver high-quality data for tens of thousands of solar-like oscillators. Future efforts should therefore extend this work to more complex internal dynamics and to other classes of oscillation modes. Indeed, theoretical studies predict that magnetic fields both reduce the excitation of $p$ mode and enhance their damping \citep{Bessila2024a}, while the energy flux of gravity and gravito-inertial modes is also predicted to depend on rotation rate \citep{Mathis2014, Augustson2020}. We plan to investigate these effects in forthcoming studies.
Overall, this work illustrates the role of fully compressible simulations in supporting asteroseismic missions by providing physically grounded constraints on mode properties and thereby maximizing the scientific outcomes from observed power spectra. 

\section*{Acknowledgements}
The authors acknowledge the reviewer, Pr. Adrian Barker, for his detailed report and constructive comments, which have helped to improve the manuscript.
A.L.S would like to thank Pr. A. S. Brun for interesting and fruitful discussions on simulations of rotating convection, as well as Dr. T. Guillet for help with the MUSIC code. L.B thanks Dr. D. Lecoanet for fruitful discussions.
The authors acknowledge support from the European Research Council (ERC) under the Horizon Europe programme (Synergy Grant agreement 101071505: 4D-STAR) from the CNES SOHO-GOLF and PLATO grants at CEA-DAp, and from ATPS (CNRS/INSU). While partially funded by the European Union, views and opinions expressed are however those of the author only and do not necessarily reflect those of the European Union or the European Research Council. Neither the European Union nor the granting authority can be held responsible for them.
This project was provided with computing HPC and storage resources by GENCI at CINES thanks to the grant 2025-A0180416177 on the supercomputer Adastra's GENOA partition. 

\section*{Data Availability}

Numerical scripts for postprocessing will be made available on a Zenodo repository at XXX if the manuscript is accepted.
The authors are happy to share the underlying spectral data of this work on a reasonable request.



\bibliographystyle{mnras}
\bibliography{biblio} 




\appendix

\section{Numerical setup}
\label{app:sims}

\subsection{The \textsc{MUSIC} code}
The five simulations presented in this study have been run using the stellar hydrodynamics code \textsc{music} \citep{Viallet2016, Goffrey17}, which solves the fully compressible equations of fluid dynamics for an invicid fluid using finite volume and time implicit methods \citep{Viallet2013}. These equations for mass, momentum and internal energy conservations are written:
\begin{equation}
\frac{\partial \rho}{\partial t} = - \vec{\nabla} \cdot (\rho \vec \velvec),
\end{equation}

\begin{equation}
\frac{\partial(\rho \vec\velvec)}{\partial t} = - \nabla \cdot (\rho \vec \velvec \otimes \vec \velvec) - \vec{\nabla}p + \rho\vec{g} - 2 \rho\vec{\Omega} \times \velvec ~, 
\end{equation}

\begin{equation}
\frac{\partial (\rho e)}{\partial t} = - \vec{\nabla} \cdot (\rho e \vec\velvec) - p \vec{\nabla} \cdot \vec \velvec - \vec{\nabla} \cdot \vec{F_r} + \rho \epsilon_{\rm nuc},
\end{equation}
where $\rho$ is the density, $e$ the specific internal energy, $\vec \velvec$ the velocity field in the rotating frame, $p$ the gas pressure, $\Omega$ is the rotational angular frequency, $\epsilon_{\rm nuc}$ is the specific energy released by nuclear burning and $\vec{g}$ the gravitational acceleration. 
The hydrodynamical simulation run for this work assumes spherically symmetric gravitational acceleration, $\vec{g} = - g \vec {\rm e}_r$, which is updated after each time step:
\begin{equation}
    g(r) = 4\pi \frac{G}{r^2} \int^r_0 \overline{\rho}(u)u^2 {\rm d}u,
\end{equation}
with $\overline{\rho}(r)$ the radial density profile given by
\begin{equation}
    \overline{\rho}(r) = \AvgS{\rho}
\end{equation}
where the operator $\AvgS{.}$ is an angular average over the whole unit sphere, defined as
\begin{equation}
    \left< h \right>_{\mathcal{S}} \equiv \frac{1}{4 \pi} \int_\mathcal{S} h(\theta, \varphi)  \sin \theta {\rm d} \theta {\rm d} \varphi.
    \label{eq:angular_av}
\end{equation}

For the stellar simulations considered here, the major heat transport that contributes to thermal conductivity is radiative transfer characterised by the radiative flux  $\vec{F_r}$, given within the diffusion approximation by

\begin{equation}
\vec{F_r} = - \frac{16 \sigma T^3}{3\kappa \rho} \vec{\nabla} T = - \chi \vec{\nabla} T,
\label{eq:radiative_flux}
\end{equation}
where $\kappa$ denotes the Rosseland mean opacity of the gas, $\sigma$ the Stefan–Boltzmann constant and $\chi$ the thermal conductivity respectively.
\textsc{music} incorporates realistic stellar opacities \citep[OPAL, ][]{Iglesias1996} for solar metallicity and equations of state \citep[OPAL EOS, ][]{Rogers2002} suitable for describing stellar interiors.

\subsection{Numerical domain and boundary conditions}
For the five simulations, we use a uniform grid resolution of
$N_r$ × $N_{\theta}$ = 1024 $\times$ 768 cells. Concerning the radial boundary conditions, we impose a constant radial derivative of the density on the inner and outer radial boundaries, as discussed in \citet{Pratt2016}. For the velocity, we impose reflective conditions at the radial boundaries. The energy flux at the inner and outer radial boundaries is set to the value of the energy flux at that radius in the initial 1D model. At the boundaries in the latitudinal direction, we impose reflective boundary conditions for the density, energy and the radial and latitudinal component of the velocity. For the azimuthal component of the velocity, we impose $\vel_{\phi}$ = 0 in $\theta = 0$ and $\pi$. 
Each simulation is initialised with a uniform reference rotation rate $\Omega$ which is kept constant during the runs. Our five simulations are summarized in Table \ref{tab:simulations_summary}. We define the convective turnover time $\tau_{\rm conv}$ by
\begin{equation}
\tau_{\rm conv} \equiv  \left< \int_{r_{\rm conv}}^{r_{\rm out}} \frac{{\rm d} r} { \vel_{\rm avg}(r,t)  } \right>_t ,
\label{eq:tau}
\end{equation}
where $r_{\rm conv} = 0.722 \Rstar$ is the location of the convective boundary of the stellar model as defined by the Schwarzschild criterion and $r_{\rm out}= 0.98 \Rstar$ is the outer boundary of our simulations. The operator $\AvgT{.}$ denotes a temporal average, and it is defined by
\begin{equation}
    \left< f(t) \right>_t \equiv \frac{1}{(t_1 - t_0)} \int_{t_0}^{t_1} f(t) {\rm d} t,
     \label{eq:avgT}
\end{equation} 
with [$t_0$, $t_1$] the time interval considered for the average.
The angular average of the velocity $\vel_{\rm avg}$ is defined by 
\begin{equation}
    \vel^2_{\rm avg}(r, t) = \sum_i \AvgS{\vel_i^2(r,\theta,t)},
    \label{eq:vavg}
\end{equation}
with $i$ = \{$r$, $\theta$\}. 

\begin{table}
\caption{\label{tab:simulations_summary}Summary of the simulations properties. With $\tau_{\rm conv}$ the convective turnover time, T$_{\rm sim}$ the total physical run time of the simulations and $N_{\rm conv}$ the number of convective turnover time covered by the simulations.}
\centering
\begin{tabular}{lcccc}
\hline\hline
$\Omega / \Omega_{\odot}$&$\tau_{\rm conv}$ ~[s]&T$_{\rm sim}$~ [s]&$N_{\rm conv}$\\
\hline
0       &4.32998$\times 10^5$ &1.5593$\times 10^8$& 360\\
1       &3.90346$\times 10^5$ &1.9450$\times 10^8$& 498\\
3       &4.12252$\times 10^5$ &2.6001$\times 10^8$& 631\\
5       &4.52976$\times 10^5$ &2.9848$\times 10^8$& 659\\
8       &5.44011$\times 10^5$ &3.1485$\times 10^8$&  579\\
\hline
\end{tabular}
\label{tab:simulations_prop}
\end{table}

Regarding spatial resolution, we have run an additional simulation at solar rotation rate with a double resolution, i.e. $N_r \times N_{\theta}$ = 2048 x 1536. This simulation has only run for 2.0 $\times 10^7$ s, but from preliminary results we have observed the usual impact of increased resolution in hydrodynamical simulations with implicit viscosity, which mostly consists in slightly higher convective velocity due to lower numerical dissipation. However, the structure of the flows does not seem impacted by the higher resolution. These results are consistent with the study of \citet{Pratt2016}, who invistaged in more details the impact of spatial resolution on convective flows in the \textsc{MUSIC} code. As our five simulations are identical (resolution, radial extent, stellar structure, etc.) except for the rotation rate, we expect that this increase in convective velocities link to smaller resolution would be very similar in all simulations. Consequently, the acoustic modes excited by convection might have slightly larger amplitudes in simulations with smaller resolution. As we are interested here in the modulation of the velocities by rotation, i.e. the ratio of the amplitudes in a rotating case to the non-rotating case, we expect that the impact on the amplitudes would be similar in all simulations, with no significant impact on the computed amplitude ratios.

\subsection{Time integration}
Following convergence to steady state, we employ the second-order time-integration scheme TR-BDF2 \citep{Hosea1996} to minimize numerical damping effects. The maximum hydrodynamic Courant number, which is related to the sound speed, is constrained to 100, while the maximum advective Courant number, relative to the flow velocity, is limited to 0.1. With this setup, which is the same for all simulations, we expect that the numerical damping due to the time integrator to be sufficiently small in the frequency range of interest for our study, which corresponds to waves with frequency smaller than 4 mHz.

\subsection{Initial 1D model}
The multidimensional simulations are initialised with the radial profiles of density and internal energy of a 1D stellar evolution model. For this work, we use a 1D model of the main-sequence 1 M$_{\odot}$ star used in \citet{Bessila2024}. This model was build using the \textsc{MESA} stellar evolution code \citep{Paxton2011, Paxton2013, paxton2015, paxton2018, paxton2019, jermyn2023}. It uses the same opacities and equation of state as implemented in \textsc{MUSIC}. The main properties of this 1D model are reported in Table \ref{tab:1D_prop}.

\begin{table}
\caption{\label{tab:1D_prop}Properties of the initial 1D stellar model used for the \textsc{MUSIC} simulations: stellar mass, luminosity, effective temperature, metallicity, radius of the convective envelope (corresponding to the location of the Schwarzschild boundary) and the pressure scale height at the convective boundary. We use $L_{\odot} = 3.8275\times 10^{33}$ erg.s$^{-1}$, $R_{\rm star} = $ 7.419 $\times 10^{10}$ cm as referenced values.}
\centering
\begin{tabular}{lccccccc}
\hline\hline
$M/M_{\odot}$&$L/L_{\odot}$&T$_{\rm eff}$ [K]&Z&$R_{\rm conv}/R_{\rm star}$&$H_{P,\rm CB}$ [cm]\\
\hline
1       &1.126 &5758&0.02& 0.722&1.914 $\times 10^{8}$\\
\hline
\end{tabular}
\label{tab:simulations_prop}
\end{table}

\section{Root-mean-square velocity}
\label{app:rms}
The rms velocity is computed using 
\begin{equation}
    \vel^2_{\rm rms}(r) = \sum_i \AvgST{\vel_i^2(r,\theta,t)} ~,
    \label{eq:vrms}
\end{equation}
with $i$ the component of the velocity considered for the computation. The operator $\AvgS{.}$ is the angular average over the whole unit sphere defined by Eq. \eqref{eq:angular_av} and $\AvgT{.}$ the temporal average defined by Eq. \eqref{eq:avgT}.

\begin{figure}
   \resizebox{\hsize}{!}
            {\includegraphics[]{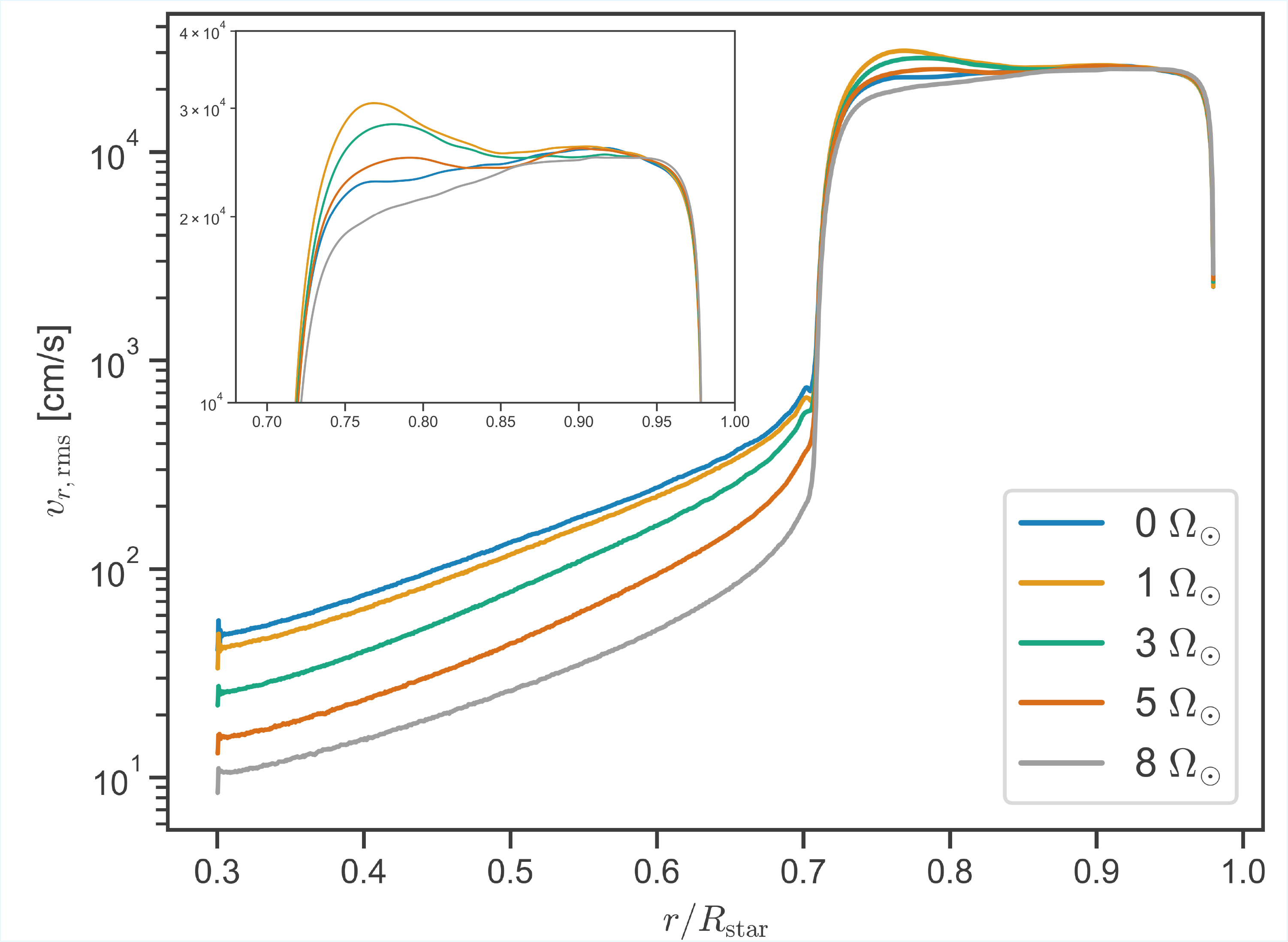}}
      \caption{Root-mean-square velocity computed using the radial velocity component. The inset is a zoom around the convective region.
              }
         \label{fig:vrms}
\end{figure}

\begin{figure}
   \centering
   \resizebox{0.9\hsize}{!}
            {\includegraphics[]{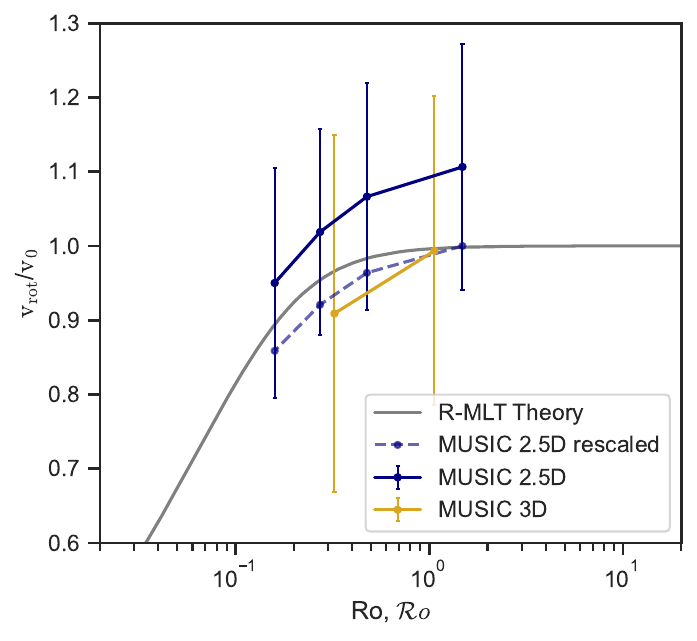}}
      \caption{Modulation of the convective velocity predicted by R-MLT (grey curve) and measured in the 2.5D (solid blue curve) and 3D (solid yellow curve) simulations as a function of the Rossby number. For these two curves, errors bars estimate the fluctuations of $\vel_{r,{\rm rms}}$ around its mean value accross the convective region. The maximal value of the velocity modulation rescaled to 1 (dashed blue curve) shows a better agreement with theory.
              }
         \label{fig:RMLT}
\end{figure}

Figure \ref{fig:vrms} shows the radial profiles of the rms radial velocity in the simulations calculated using Eq. \eqref{eq:vrms}. 
This illustrates the impact of rotation on convective velocities in the radial direction, which is the quantity of interest when comparing to the 1D Mixing Length Theory (MLT). Looking at the rotating models, it appears that as the rotation rate increases, the rms velocities decrease in the convection zone. This decrease is particularly obvious at the bottom of the convection zone, near the convective boundary, as illustrated in the insert of Fig. \ref{fig:vrms}. This behaviour is predicted by the Rotating MLT \citep{Stevenson1979, Augustson2019, Bessila2025}, as illustrated in Fig. \ref{fig:RMLT}. Note that the Rossby numbers computed from the simulations and for the theoretical model are a bit different, and are defined using Eq. \eqref{eq:rossby} and Eq. \eqref{eq:rossby_mlt}, respectively. We refer the reader to Appendix \ref{app:rossby} for more details.
However, this trend does not seem to hold in the non-rotating case, which should have the fastest velocities. This is why the ratio of the velocities in Fig. \ref{fig:RMLT} are larger than one for the \textsc{MUSIC} curve (plain blue). We can see that if we rescale it to one (dashed blue curve), the agreement with R-MLT is even better. The 2D geometry is known to constraint convective flow differently than 3D \citep[see for exemple][ and references therein]{Pratt2020}, which could explain this discrepancy. In order to test this hypothesis, we are currently running 3D simulations of the same stellar model. From preliminary results, it seems that the 3D simulations tend to show a better agreement with R-MLT. Indeed, as shown in Fig. \ref{fig:RMLT}, velocities in the non-rotating case are larger than in the 1 and 3 $\Omega{\odot}$ cases in the 3D simulations.

\section{Rossby number}
\label{app:rossby}
In hydrodynamical simulations it is usual to define the Rossby number $\mathrm{Ro}$ as in \citet{Hindman2020}, i.e.
\begin{equation}
    \mathrm{Ro} = \frac{\vel_{\rm rms}}{2 \Omega_0 H},
    \label{eq:rossby}
\end{equation}
with $\Omega_0$ the rotation rate imposed in the simulation, $\vel_{\rm rms}$ the root-mean-square velocity, which is defined in Eq. \eqref{eq:vrms}, and $H$ the depth of the convective shell, which extend here from $r_{\rm conv} = 0.722$ to $r_{\rm out} = 0.98 \Rstar$. The Rossby number represents the relative importance of the advection to the Coriolis force. In the regions where the motions are aligned with convection, we measure a Rossby number less than one, ${\rm Ro} \lesssim 0.3$, for the 8$\Omega_{\odot}$ cases. It means that the Coriolis force dominates over advection and thus explains the presence of Taylor columns. In the region close to the equator we have $Ro$ $\sim 1$ confirming that advection is stronger in this region. In the solar rotating case, there is $Ro$ $\gtrsim 1$ everywhere in the convective zone, therefore convective motions are only weakly impacted by rotation. 

In the context of the R-MLT, the Rossby number is computed as a function of the non-rotating characteristic velocity $\vel_0$ and the convective scale $k_0$ \citep{Stevenson1979}
\begin{equation}
    \mathcal{R}o = \frac{\vel_0k_0}{2 \Omega_0 \cos \theta},
    \label{eq:rossby_mlt}
\end{equation}
with $\theta$ the co-latitude. See Eq. (47) of \citet{Bessila2024} for details.

\section{Amplitudes of acoustic modes up to $\ell = 10$ and 20}
\label{app:extra_ampl}

\begin{figure}
   \resizebox{\hsize}{!}
            {\includegraphics[]{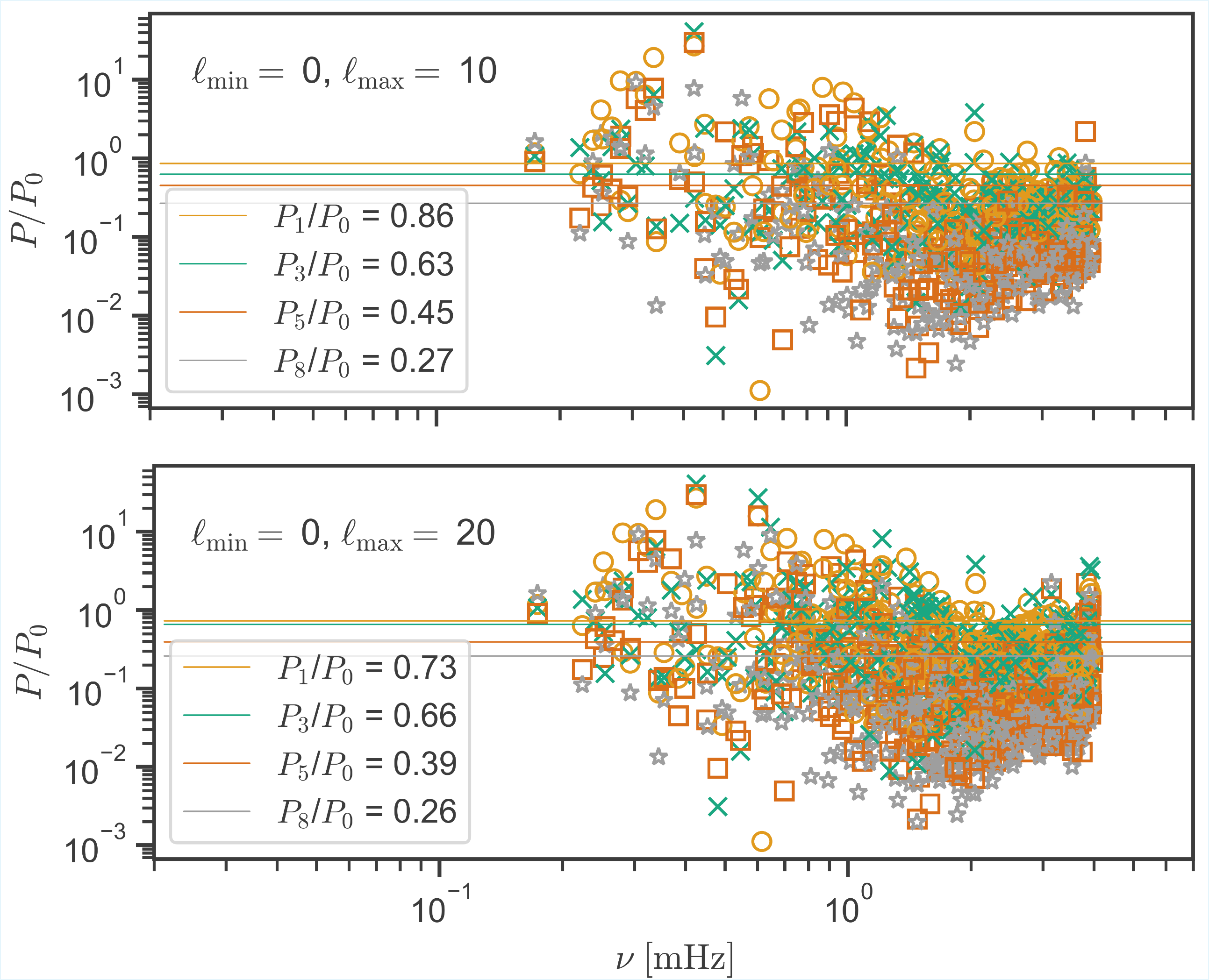}}
      \caption{Ratios of the  $p$ modes' amplitudes in a rotating simulation to the non-rotating case as a function of frequency for the 5 studied rotation rates for angular degrees $\ell$ between 0 and 10 (upper panel) and between 0 and 20 (lower panel). The amplitudes of the modes are computed at r = 0.9 R$_{\rm star}$. The horizontal lines represent the amplitude ratios averaged over all modes of a given simulation.
              }
         \label{fig:comp_largeL}
\end{figure}

Figure \ref{fig:comp_largeL} shows the same diagnostic as in Fig. \ref{fig:comp_ellrange} but including $p$ modes of larger degrees. The upper panel includes $p$ modes with $\ell \in [0,10]$ and the lower panel the ones with $\ell \in [0,20]$. For both panels we use modes with frequencies $\nu$ between the Lamb frequency $L_{\ell}$ and 4 mHz. The ratios of the amplitude change slightly depending on the angular degrees included, but the main behaviour remains similar with a clear decrease of the amplitudes with increasing rotation rate. We report the values of the decreased amplitudes, as well as damping rates, for the different intervals of angular degrees in Table \ref{tab:decrease}.

\begin{table}
\caption{\label{tab:decrease} Reduction in amplitude $\Delta A$ and damping rate $\Delta \eta$ of $p$ modes in rotating simulations with respect to the non-rotating simulation for different intervals of angular degrees $\ell \in $ [0, $\ell_{\mathrm{max}}$]. Negative values of decrease corresponds to an incrase in damping rate.}
\centering
\begin{tabular}{lccc}
\hline\hline
$\Omega / \Omega_{\odot}$&$\ell_{\mathrm{max}}$&$\Delta A$ [\%] &$\Delta \eta$ [\%]\\
\hline
1       & 5 & 21 & 3\\
1       & 10 & 14 & 4\\
1       & 20 & 27 & 7\\
\hline
3       & 5 & 29 & -15\\
3       & 10 & 37 & -21\\
3       & 20 & 34 & -21 \\
\hline
5       & 5 & 53 & 24\\
5       & 10 & 55 & 18\\
5       & 20 & 61 & 28\\
\hline
8       & 5 & 77 & 51\\
8       & 10 & 73 & 54\\
8       & 20 & 74 & 60\\
\hline
\end{tabular}
\label{tab:simulations_prop}
\end{table}

\section{Eddy-time correlation function}
\label{app:eddy}
A crucial hypothesis of the work of \citet{Bessila2024} and \citet{Biscarrat2025} is to assume that the convective eddies are temporally correlated following a Lorentzian function. 
As described in Sect. \ref{sec:theory}, this correlation is modelled through the eddy-time correlation function $\chi_k$, following the standard formalism adopted in models of stochastic excitation of gravity and acoustic modes \citep[e.g.][]{Goldreich1994, Samadi2001, Belkacem2010}. Initial works considered a Gaussian correlation function of the form \citep{Goldreich1977, Balmforth1992, Samadi2001}:
\begin{equation}
    \chi^G_{k}(\omega) = \frac{1}{\sqrt{\pi}\omega_k}\exp\left( -\frac{\omega^2}{2\omega_k^2} \right),
    \label{eq:chi_g}
\end{equation}
but more recent studies showed that a Lorentzian function:
\begin{equation}
    \chi^L_{k}(\omega) = \frac{1}{\pi\omega_k} \frac{1}{1+\left( \omega^2/\omega_k^2 \right)},
    \label{eq:chi_l}
\end{equation}
agrees better with solar observations and numerical simulations \citep{Samadi2003,Belkacem2010,Philidet2023}.

In Eq. \eqref{eq:chi_g} and \eqref{eq:chi_l}, the linewidth $\omega_k$, which is also the frequency associated with the eddy of wavenumber $k$, is defined as \citep{Belkacem2009}
\begin{equation}
    \omega_k \equiv \frac{2 k u_k}{\lambda},
    \label{eq:omegak}
\end{equation}
with $\lambda$ a parameter as in \citet{Balmforth1992} and $u_k$ the velocity of the eddy with wavenumber $k$. As described in \citet{Stein1967}, this velocity can be related to the kinetic energy spectrum $E(k)$ with
\begin{equation}
    u_k^2 = \int^{2k}_k E(k) \mathrm{d}k.
\end{equation}
To estimate $u_k$ from the simulations used in this work, we use the same method as in \citet{Belkacem2009} and calculate it from the power spectrum of the velocities computed in the convection zone of our simulations. The kinetic energy spectrum $E(\ell)$ is defined following \citet{Samadi2003} as
\begin{equation}
    E(r,\ell) = \sum_p \AvgT{\left(\tilde{\vel}_{p}-\AvgT{\tilde{\vel}_p}\right)^2}
    \label{eq:Ek_l}
\end{equation}
with $\tilde{\vel}_p$ the component $p = \{r, \theta\}$ of the velocity decomposed on the basis of spherical harmonics and $\AvgT{.}$ is a time average as defined in Eq. \eqref{eq:avgT}. As in \citet{Belkacem2009}, we do not include the density in the kinetic energy spectrum in order to compare with their results. The kinetic energy spectrum $E(\ell) = E(r_0,\ell)$ measured in the non-rotating simulation at $r=0.9 R_{\rm star}$ is shown in Fig. \ref{fig:KEspec}. The top panel shows the kinetic energy spectrum computed using $\vel_r$ and $\vel_{\theta}$, the radial and latitudinal components of the velocity respectively, and the bottom panel using only $\vel_r$. The two power spectra are similar at large angular degrees, but differ at low degrees. There is much less power in large scales, i.e. low $\ell$, in radial motions. One possible explanation is that this difference is related to mean flows in the simulations.
If we examine the overall shape of the power spectra, we see that they are not well represented by a power-law distribution, although this may be the case locally. In addition, this shape is very similar to the one obtained by \citet{Belkacem2009} and \citet{Alvan2014} for their kinetic energy spectra computed with the same method. We find that the kinetic energy spectra are maximum for $\ell$ =  21, 30, 24, 33 and 42 in the $\Omega =$ 0, 1, 3, 5 and 8 $\Omega_{\odot}$ case, respectively. By estimating the horizontal length scale of convective eddies as $l_{\rm h} \sim k_{\rm h}^{-1}$, with $k_{\rm h} = \sqrt{\ell(\ell+1)}/r$, our results suggest that larger rotation rate corresponds to smaller horizontal convective length scale. This is coherent with the decrease of the characteristic horizontal length scale of convection with increasing rotation rate predicted by R-MLT \citep{Stevenson1979, Bessila2025} and measured in simulations \citep{Brown2008, deVries2023}. This decrease is related to the deformation of convective eddies by the Coriolis force.

We also introduce the kinetic energy spectrum as a function of the frequency $\nu$ 
\begin{equation}
    E(r,\ell,\nu) = \sum_p P[\hat{\rm v}_p^2]
    \label{eq:Ek_L_nu}
\end{equation}
with $P[\hat{\rm v}_p^2]$ the power spectrum the of component $p$ of the velocity as introduced in Sect. \ref{sec:pmodes_ampl}.
Then, as in \citet{Samadi2003a} and \citet{Belkacem2009}, we decompose $E(r,\ell,\nu) $ as
\begin{equation}
    E(r,\ell,\nu) = E(r,\ell)\chi_{\ell}(r,\nu) ,
    \label{eq:chik}
\end{equation}
with $\chi_{\ell}(r,\nu)$ the eddy time correlation that satisfies
\begin{equation}
   \int \chi_{\ell}(r,\nu) \rm{d}\nu =1.
\end{equation}

\begin{figure*}
   \resizebox{\hsize}{!}
            {\includegraphics[]{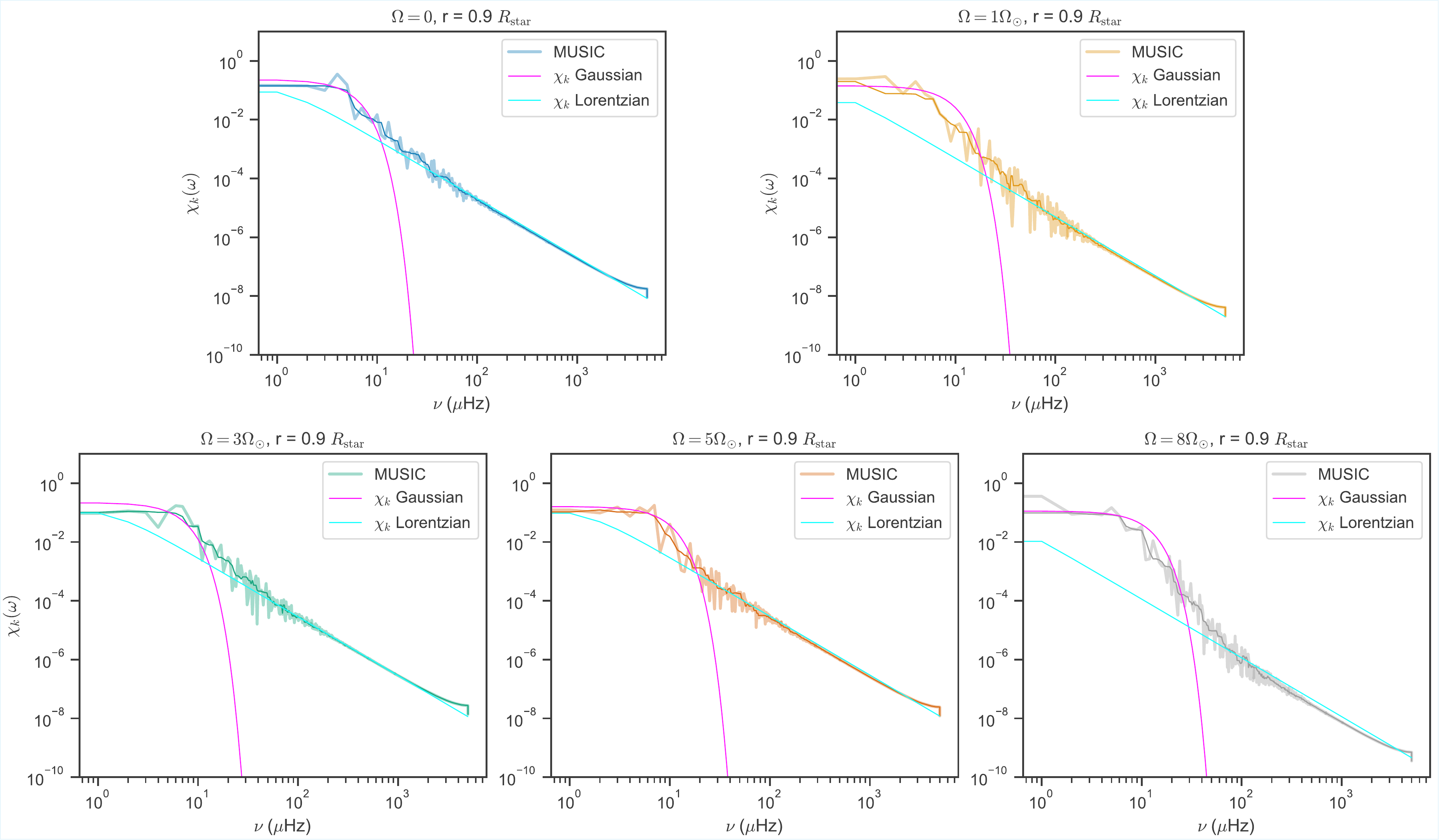}}
      \caption{Eddy-time correlation function $\chi_k$ measured at r = 0.9 $R_{\rm star}$ in the five simulations, plotted as the thick line in each plot. The thin line of same color is a smoothing average performed overs 11 frequency bins for better comparison with analytical functions, which are a Gaussian (magenta line) and a Lorentzian (cyan) functions.
              }
         \label{fig:chi_k}
\end{figure*}

\begin{figure}
   \resizebox{\hsize}{!}
            {\includegraphics[]{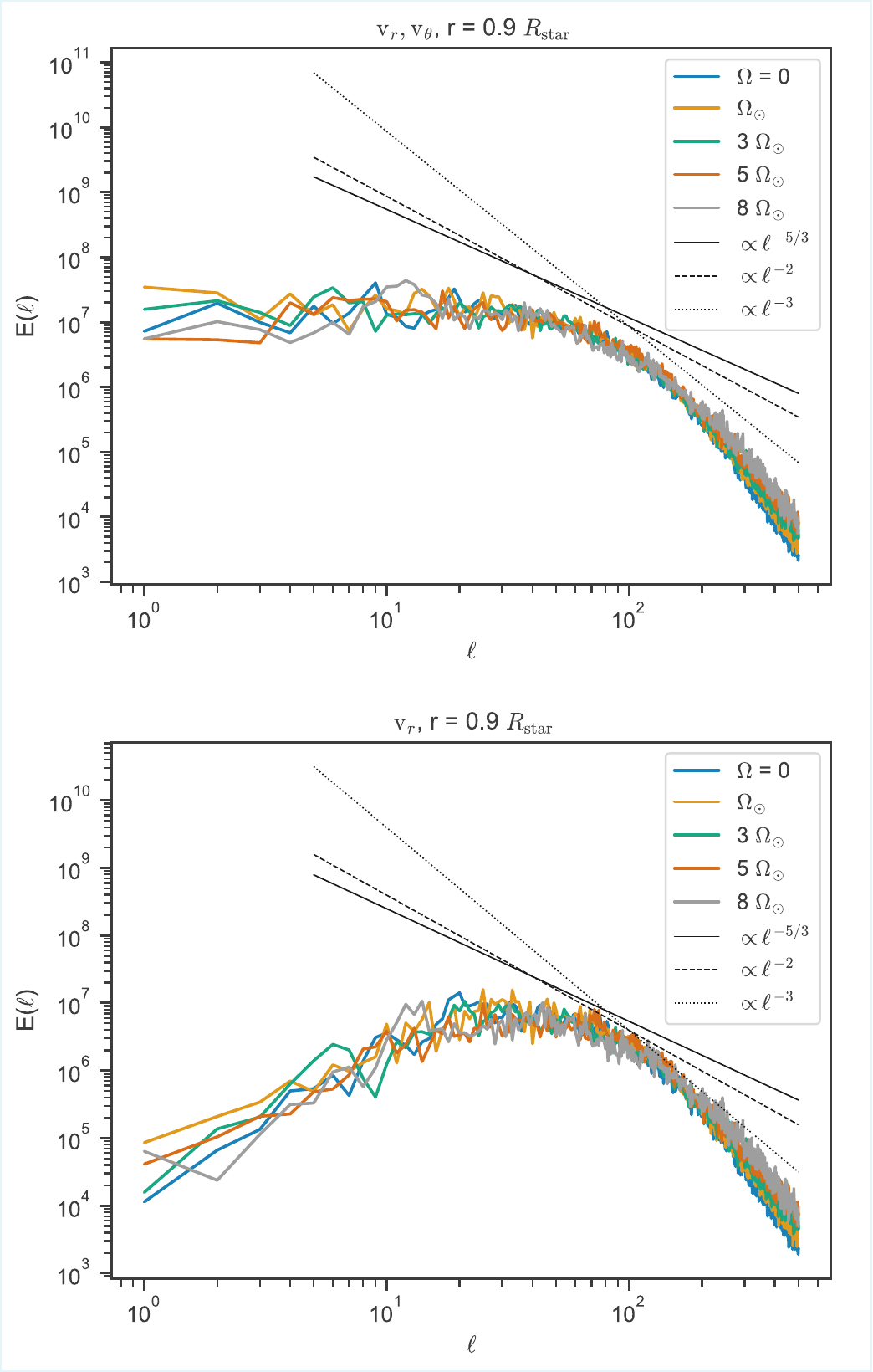}}
      \caption{Kinetic energy spectrum $E(\ell)$ measured using the radial and latitudinal component of the velocity (top panel) and using only the radial component (bottom) at r = 0.9 $R_{\rm star}$ in the five simulations.
              }
         \label{fig:KEspec}
\end{figure}

Figure \ref{fig:chi_k} compares the eddy-time correlation function $\chi_k$ measured in the simulations, computed using Eq. \eqref{eq:Ek_l} - \eqref{eq:chik}, to the theoretical correlation functions $\chi_k^L$ (cyan) and $\chi_k^G$ (magenta) computed using Eq. \eqref{eq:chi_l} and \eqref{eq:chi_g}, respectively. For these two theoretical functions, the frequency associated with an eddy with wavenumber $k$ is computed using Eq. \eqref{eq:omegak} where $u_k$ is estimated from the simulation using Eq. \eqref{eq:Ek_l}. In addition, the free parameter $\lambda$ in Eq. \eqref{eq:omegak} is adjusted to offer the best agreement with the simulations. We use $\lambda$ = 3, 10, 3, 4, and 24 for the Lorentzian function in the $\Omega =$ 0, 1, 3, 5 and 8 $\Omega_{\odot}$ case respectively, and $\lambda$ = 1/2 for the Gaussian one for all simulations.

We found that the eddy-time correlation function is far better approximated using a Lorentzian function in the $p$ modes frequency range for all rotation rates considered in this study. This confirms the hypothesis used by \citet{Bessila2024}. In addition, this result agrees with previous estimations of this correlation function from numerical simulations of the solar interior and observations \citep{Belkacem2009, Alvan2014, Philidet2023}.

\section{Mode damping}
\label{apdx:damping}
\begin{figure}
   \resizebox{\hsize}{!}
            {\includegraphics[]{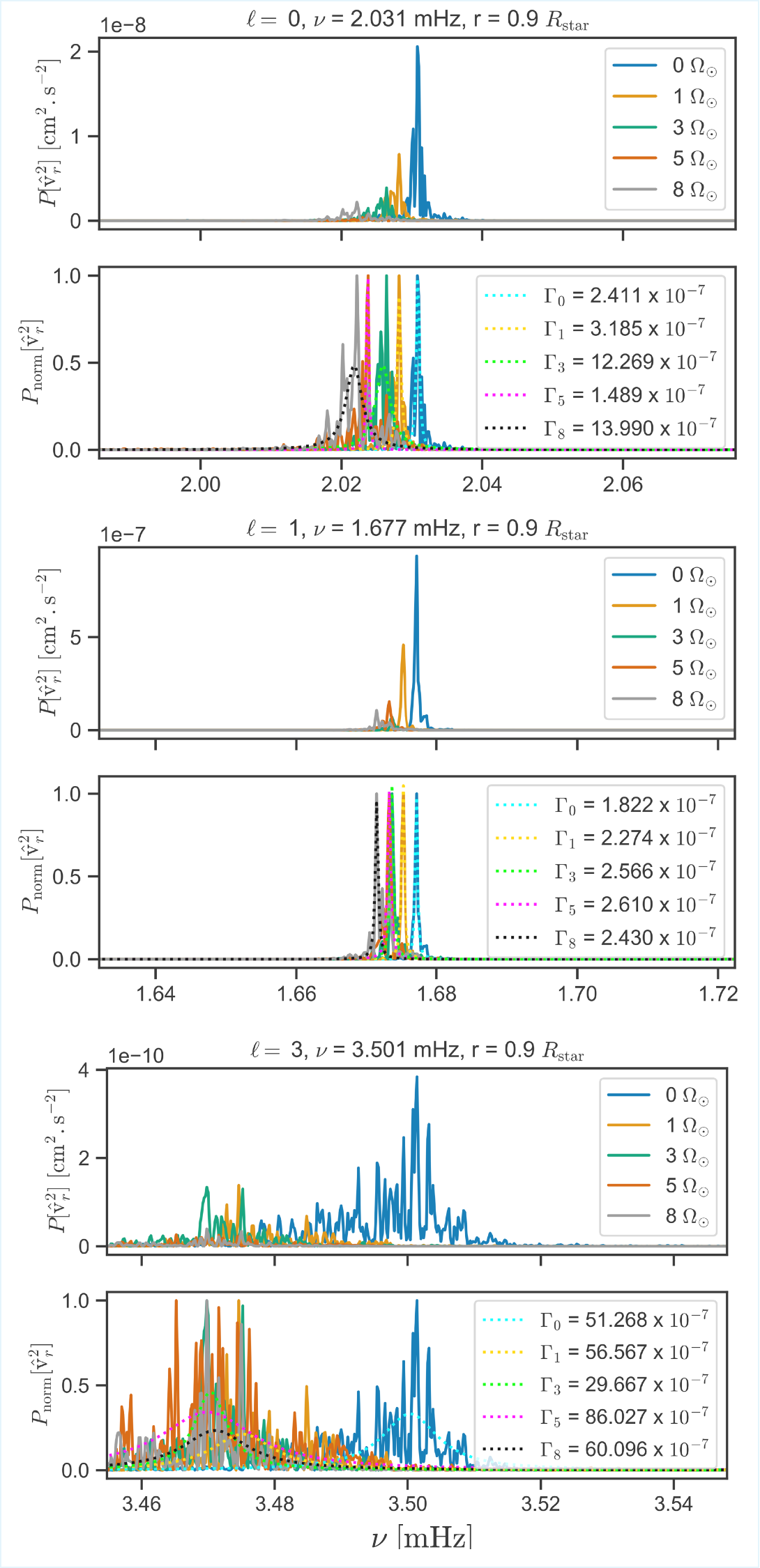}}
      \caption{Measurement of the half linewidth $\Gamma$ of three $p$ modes $(\ell, \nu) = (1, 0.9616)$ (top), (5, 1.587) (middle) and (2, 2.227) (bottom). For each mode the upper plot shows the power spectra computed using the radial velocity in each simulation centred around the frequency of the mode, and the lower plot shows the normalised power spectrum (plain curves) and the Lorentzian fit (dotted lines).
              }
         \label{fig:damping}
\end{figure}

The amplitude of solar-like oscillations is set by the competing effects of growth due to power injection by turbulent convection and decay due to damping by convective and radiative effects. In simulations, the power spectrum of the radial velocity is an estimation of the squared amplitude of a mode, thus we have $P[\hat{\rm v}_r^2] \sim A^2$.  The amplitude we measure in the simulations can be approximated as
\begin{equation}
    P[\hat{\rm v}_r^2] \propto A^2 \propto \frac{\mathcal{P}}{\eta},
    \label{eq:Pinj_sims}
\end{equation}
with $A$ the amplitude and $\eta$ the damping rate of the considered mode and $\mathcal{P}$ the power injected into the mode, as defined Eq.\eqref{eq:power}. 
As mentioned in the Sect. \ref{sec:mode_damping}, the damping rate $\eta$ of a mode can be directly inferred from the simulations by measuring the linewidth of the corresponding peak in the power spectrum of the radial velocity.
First, we isolate a peak in the power spectrum by selecting $\sim$ 450 frequency bins around the central frequency, which represents 90 $\mu$Hz or half the frequency interval between two consecutive $p$ modes. We then normalise its value to 1 to get $P_{\rm norm}[\hat{\rm v}_r^2]$, which is shown on the bottom panel of the plots corresponding to each mode in Fig. \ref{fig:damping}.

Then, using the \textit{curve fit} function of the \textsc{SciPy} python library \citep{SciPy2020}, we fit the mode with a Lorentzian function $f_L$ of the form :
\begin{equation}
    f_L(I,\Gamma,\nu) = \frac{I\Gamma^2}{\Gamma^2 + ({\nu - \nu_0)^2}},
\end{equation}
with $I$ the mode height and $\nu_0$ the eigenfrequency of the mode predicted by GYRE.
It gives a measurement of the half width at half maximum (HWHM) $\Gamma$. This is illustrated in Fig. \ref{fig:damping_l1_main} for the mode identified with $(\ell, \nu) = (4, 2.432)$ and in Fig. \ref{fig:damping} for three different $p$ modes identified with $(\ell, \nu) = (0, 2.031)$ (top panel), (1, 1.677) (middle) and (3, 3.501) (bottom), where $\nu$ is expressed in mHz. This quantity can be linked to the modes' damping rate with the relation given in Eq. \eqref{eq:damping}. We draw attention to the fact that this method has a significant limitation related to finite duration of the analysed time series. To compute the Fourier transforms, we apply a Blackman window to apodise the non-periodic signals \citep[see][]{LeSaux2022}. Using a time span of 5 $\times 10^6$ s to compute the power spectra, we obtain a frequency resolution of 2 $\times 10^{-7}$ Hz. Consequently, mode linewidth smaller than 2 $\times 10^{-7}$ Hz, i.e. $\Gamma$ < 1 $\times 10^{-7}$ Hz, cannot be reliably resolved. This means that modes with instrinsically narrow peaks and thus long lifetimes, i.e. very small damping rates, are dominated by the Blackman window function. Those are primarily the lowest frequency $p$ modes \citep{Davies2014}. 
To ensure that only guenuinely damped modes are retained, we restrict our analysis to modes with $\Gamma$ > 4 $\times 10^{-7}$ Hz. For instance, the mode $(\ell, \nu) = (1, 1.677)$ shown in the middle panel of Fig. \ref{fig:damping} is disregarded in all simulations. 
Modes with unrealistically large fitted linewidths ($\Gamma > 300 \times 10^{-7}$ Hz), typically arising from failed Lorentzian fits at high frequency where signal-to-noise ratios are low, are also excluded.

The measured values of individual damping rates can vary significantly from one mode to another. Therefore, as for the amplitude of the modes in Fig. \ref{fig:comp_ellrange}, we average the damping rates over a large set of $p$ modes in each simulation. 
Using Eq. \eqref{eq:damping}, we obtain mean damping rates $\eta_0 = 26.84$ $\mu$Hz, $\eta_1 = 27.76$ $\mu$Hz, $\eta_3 = 22.45$  $\mu$Hz, $\eta_5 = 33.15$ $\mu$Hz and $\eta_8 = 40.59$ $\mu$Hz for the simulations with $\Omega$ = 0, 1, 3, 5 and 8 $\Omega_{\odot}$, respectively.
In \citet{Belkacem2019}, the authors report a damping rate $\eta = 157$ $\mu$Hz for a $p$ mode at 3.540 mHz. In the bottom panel of Fig. \ref{fig:damping}, we look at a mode with similar frequency and the corresponding damping rate is $\eta = 320$ $\mu$Hz, which is a factor two larger than what is found by \citet{Belkacem2019}. However, the comparison is not straightforward as in \citet{Belkacem2019} the simulation used is a local box close to stellar surface with a very different treatment of radiative heat transfer. 
Moreover, we emphasize that, as mentioned in \citet{Belkacem2019} and shown by \citet{Appourchaux2014}, linewidths determination are known to be subject to several biases. Consequently, the absolute values given here for the damping rates should be treated with caution and regarded as order-of-magnitude estimates. Our focus is therefore on relative variations with rotation. While no strictly monotonic trend is observed, the two fastest rotating models exhibit systematically larger damping rates, tentatively suggesting enhanced damping with increased rotation rate.

\section{Estimation of power injected into modes using simulations' power spectra}
\label{app:Pinj_sims_spectra}

\begin{figure}
   \resizebox{\hsize}{!}
            {\includegraphics[]{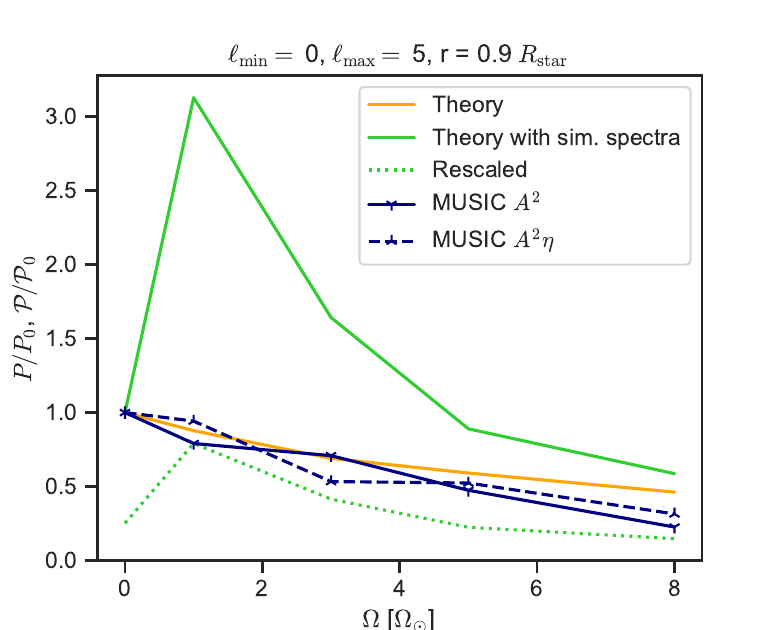}}
      \caption{Modulation with rotation rate of the ratio of $p$ modes' amplitude $P$ measured in rotating simulations to the amplitude in the non-rotating simulation $P_0$ (plain blue curves), and of the ratio of the power injected into the modes in the rotating to the non-rotating case, predicted theoretically using analytical expressions for $E(k)$ and $\chi_k(\omega)$ (grey curve), predicted theoretically using measurements in the simulations for $E(k)$ and $\chi_k(\omega)$ (cyan curves) and directly measured in simulations (dashed blue curve). The dotted cyan curve correspond to the cyan curve but rescaled to match the simulation curve (dark blue dashed line) at $\Omega = \Omega_{\odot}$.
              }
         \label{fig:Pinj_sim_spectra}
\end{figure}

As described in Sect. \ref{sec:theory}, estimating the power injected into acoustic modes by turbulent convection requires both a kinetic energy spectrum $E(k)$ and an eddy-time correlation function $\chi_k(\omega)$. \citet{Bessila2024} showed that using a Lorentzian function for $\chi_k$ is required to reproduce for the observed rotational dependence of the acoustic modes amplitude reported by \citet{Mathur2019}. In the present work, we have demonstrated that this choice is justified, as illustrated in Fig. \ref{fig:chi_k}. 
Concerning the kinetic energy spectra, \citet{Bessila2024} used a Kolmogorov power law, as it is usually the choice in analytical work studying $p$ modes excitation \citep[e.g.][]{Goldreich1994, Samadi2001}. However, as shown in Fig. \ref{fig:KEspec} the kinetic energy spectra in the convective region of our simulations deviate significantly from a simple power-law behaviour.
We therefore compute the injected power using Eq. \eqref{eq:power}, replacing analytical prescriptions with the measured $E(k)$ and $\chi_k(\omega)$ in the simulations, which are shown in Fig. \ref{fig:chi_k} and \ref{fig:KEspec}, respectively. 
In this calculation we retain only the radial velocity component and assume $k \sim k_{\rm h} = \sqrt{\ell(\ell+1)}/r$. This latter assumption is a significant difference with the full analytical calculations of \citet{Bessila2024} as in our case $k_{\rm h}$ takes values between $\sim 10^{-11}$ and $\sim 10^{-8}$, whereas $k$ takes values between $\sim 10^{-11}$ and $\sim 10^{-1}$ in full analytical treatment of \citet{Bessila2024}. The resulting power injected $\mathcal{P}$ is shown in Fig. \ref{fig:Pinj_sim_spectra} (cyan curve).

With this approach, the injected power is not maximal in the non-rotating case. It is larger by factors of $\sim$3 and $\sim$1.5 for $\Omega = 1$ and  $3 ~ \Omega_{\odot}$, comparable for $\Omega = 5 \Omega_{\odot}$, and reduced only at $\Omega = 8 \Omega_{\odot}$.
This is consistent with what we measure for the rms velocity in Appendix \ref{app:rms} and to what we understand as a bias introduced by the assumed simplified 2.5D geometry that however allows us to explore for a reasonable numerical cost the required parameter space. We do find results more consistent with R-MLT in the preliminary analysis of our 3D simulations, as illustrated in Fig. \ref{fig:RMLT}.

Importantly, when considering only rotating models, the injected power decreases systematically with increasing rotation rate. Looking at the rescaled injected power computed with measured $E(k)$ and $\chi_k(\omega)$ (dotted cyan curve), we can see that the decrease with rotation rate follow a similar decay as for the global amplitude (plain blue curve) and power injected directly measured in the simulations (dashed blue curve). It is also consistent with the theoretical predictions using analytical prescriptions for $E(k)$ and $\chi_k(\omega)$ (grey curve).

\section{Differential rotation in simulations}
\label{app:diff_rot}

\begin{figure*}
   \resizebox{\hsize}{!}
            {\includegraphics[]{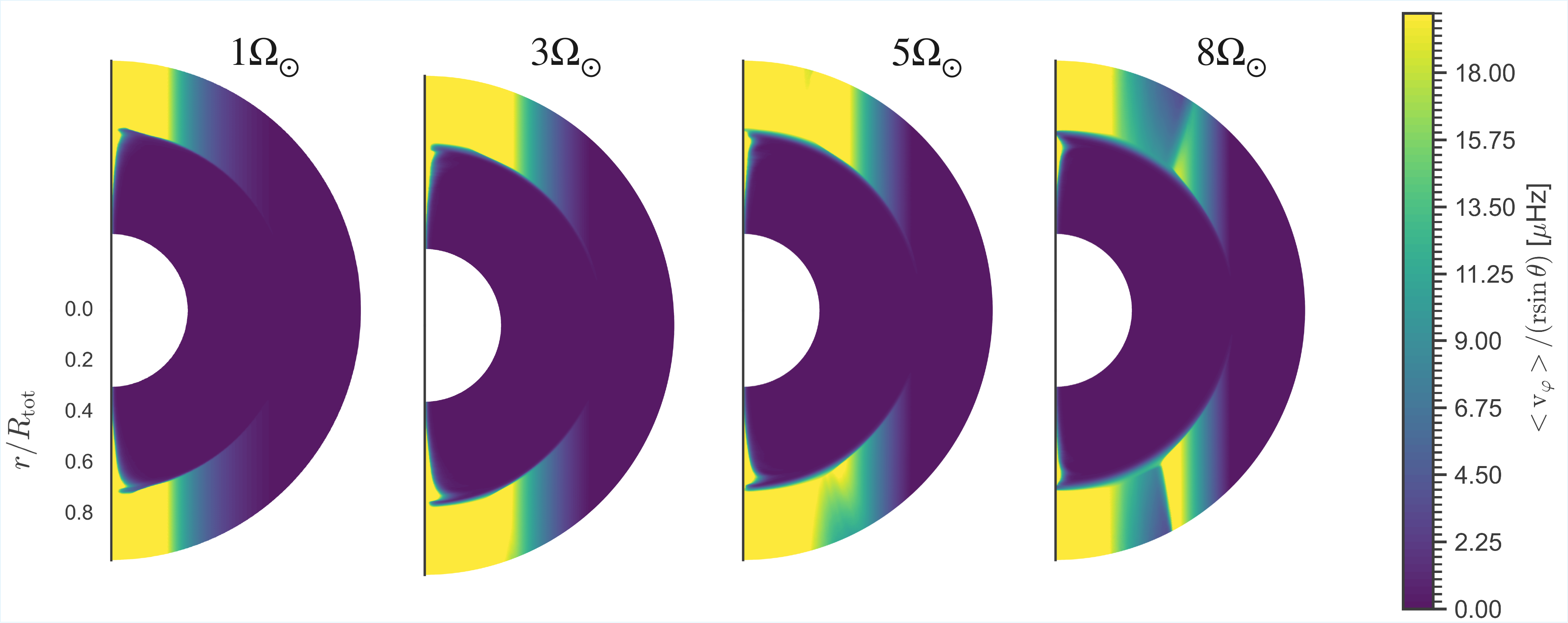}}
      \caption{Mean rotation rate in the rotating grame $\left< \vel_{\varphi}(t,r,\theta)\right>_t/(r \sin \theta )$ in the four rotating simulations.
              }
         \label{fig:diff_rot}
\end{figure*}

All the rotating simulations are initialised with a uniform rotation rate, which is kept constant during the simulations time. However, as a simulation evolve, angular momentum is redistributed. This tends to create and trigger a mean flow that can be visualised by computing the mean rotation rate $\left<\vel_{\varphi}(t,r,\theta)\right>_t/(r \sin \theta)$ for each simulation, which is presented in Fig. \ref{fig:diff_rot}. The time average is computed using Eq. \eqref{eq:avgT} over a time interval of $10^7$ s, which corresponds to approximately 20 convective turnover times (see Table \ref{tab:simulations_summary}).

For all our simulations, we can see that the poles rotates faster than the equator, which corresponds to an anti-solar differential rotation, by opposition of the solar-like differential rotation, where the equator rotates faster than the poles.
The mean rotation rate presented in Fig. \ref{fig:diff_rot} suggests a larger differential rotation as the mean rotation rate increases. We confirm this behaviour by measuring the difference in angular velocity $\Delta \Omega$ between the equator and a latitude of 60 degrees \citep{Brown2008}

\begin{equation}
    \Delta \Omega = \Omega_{\rm eq} - \Omega_{60},
\end{equation}

where $\Omega_{j}$ = $\Omega_0 + \vel_{\varphi}/(r \sin \theta_{j})$ with $j$ = \{eq, 60 \}. We measure $\Delta \Omega$ = -0.57, -1.76, -3.02, -5.17 $\mu$Hz. Despite that we observe an anti-solar-like differential rotation, the results qualitatively agree with \citet{Brown2008} who found that $\Delta \Omega$ increases with increasing rotation rate. However, the 2D geometry of our simulations might create stronger mean flow than in 3D simulations.

According to the literature, it is expected to obtain a transition from anti-solar to solar-like differential rotation for a critical Rossby number Ro $\sim 1$ \citep[see for example][]{Brun2017, Noraz2025}, which is not what we observe here. All our simulations show anti-solar differential rotation. This behaviour could be explained by the two-dimensional geometry of our simulations. It could also be linked to the Prandtl number. Indeed, as shown by \citet{Kapyla2023}, this critical Rossby number depends on the Prandtl number. A larger Prandtl tends to shift the critical Rossby towards lower values. In their study, \citet{Kapyla2023} show that even for a Prandtl number of 10, the critical Rossby number shifts from $\sim 0.7$ to $\sim 0.3$ \citep[see Fig. 5 in][]{Kapyla2023}. Thus, as we use implicit viscosity, we do not know the value of the Prandtl number in our simulations, and it is possible that the critical Rossby number is shifted to a different value. We are planning a follow-up study to analyse this in more detail.


\bsp	
\label{lastpage}
\end{document}